\documentclass[%
  reprint,
  twocolumn,
showpacs,preprintnumbers,
 amsmath,amssymb,
 aps,
 prd,
]{revtex4-1}

\usepackage{graphicx}% Include figure files
\usepackage{dcolumn}% Align table columns on decimal point
\usepackage{bm}% bold math
\usepackage{hyperref}% add hypertext capabilities
\usepackage{slashed}
\usepackage{color}
\usepackage{physics} %add some physics specific maths capabilities
\DeclareRobustCommand{\eqnr}[1]{Eq.~$\left(\ref{#1}\right)$}
\newcommand{\Fig}[1]{Fig.~\ref{#1}}
\newcommand{\Figtwo}[2]{Figs.~\ref{#1} and \ref{#2}}
\newcommand{\omk}{\omega_{\vec{k}}}

\newcommand{\upup}{\upharpoonleft\!\upharpoonright}
\newcommand{\updown}{\upharpoonleft\!\downharpoonright}

\newcommand{\vectwo}[2]{\left(\vec{#1}, #2\right)}
\newcommand{\eqnrtwo}[2]{Eqs.~$\left(\ref{#1}\right)$ and $\left(\ref{#2}\right)$}

\newcommand{\gmu}{\gamma^\mu}
\newcommand{\pmu}{\partial_\mu}
\newcommand{\Amu}{A_\mu}
\newcommand{\smunu}{\sigma^{\mu\nu}}
\newcommand{\Fmunu}{F_{\mu\nu}}

\newcommand{\vB}{\vec{B}}

\newcommand{\vs}{\vec{\sigma}}

\newcommand{\zh}{\hat{z}}

\DeclareMathOperator{\SU}{SU}

\begin{document}

\preprint{ADP-18-10/T1058}

\title{Neutron magnetic polarizability with Landau mode operators}% Force line breaks with \\

\author{Ryan Bignell}
 \email{ryan.bignell@adelaide.edu.au}
 \author{Jonathan Hall}
\author{Waseem Kamleh}%
\author{Derek Leinweber}

\affiliation{%
Special Research Centre for the Subatomic Structure of Matter (CSSM),\\
Department of Physics, University of Adelaide, Adelaide, South Australia 5005, Australia
}%

%\collaboration{CSSM Collaboration}%\noaffiliation

\author{Matthias Burkardt}
\affiliation{
 Department of Physics, New Mexico State University, Las Cruces, New Mexico 88003-001, USA
}%
%\author{Jonathan Hall}
%% \homepage{http://www.Second.institution.edu/~Charlie.Author}
%\affiliation{
%Special Research Centre for the Subatomic Structure of Matter,\\
%University of Adelaide, Adelaide, South Australia 5005 Australia
%\affiliation{%

%\collaboration{CSSM Collaboration}%\noaffiliation

             
\begin{abstract}
  The application of a uniform background magnetic field makes standard quark operators utilizing gauge-covariant Gaussian smearing inefficient at isolating the ground state nucleon at nontrivial field strengths. In the absence of QCD interactions, Landau modes govern the quark energy levels. There is evidence that residual Landau mode effects remain when the strong interaction is turned on. Here, we introduce novel quark operators constructed from the two-dimensional $U(1)$ Laplacian eigenmodes that describe the Landau levels of a charged particle on a periodic finite lattice. These eigenmode-projected quark operators provide enhanced precision for calculating nucleon energy shifts in a magnetic field. Using asymmetric source and sink operators, we are able to encapsulate the predominant effects of both the QCD and QED interactions in the interpolating fields for the neutron. The neutron magnetic polarizability is calculated using these techniques on the $32^3 \times 64$ dynamical QCD lattices provided by the PACS-CS Collaboration. In conjunction with a chiral effective-field theory analysis, we obtain a neutron magnetic polarizability of \hbox{$\beta^n = 2.05(25)(19) \cross 10^{-4}$ fm$^3$}, where the numbers in parentheses describe statistical and systematic uncertainties.
%  \\
%  \\
%DOI: \href{https://doi.org/10.1103/PhysRevD.98.034504}{10.1103/PhysRevD.98.034504}
%\begin{description}
%		\item[Usage]
%		Secondary publications and information retrieval purposes.
%		\item[PACS numbers]
%		13.40.-f, 12.38.Gc, 12.39.Fe
%		May be entered using the \verb+\pacs{#1}+ command.    %I really don't get the \pacs command -needs showpacs in document options
%		\item[Structure]
%		You may use the \texttt{description} environment to structure your abstract;
%		use the optional argument of the \verb+\item+ command to give the category of each item. 
%	\end{description}
\end{abstract}
\maketitle

%\tableofcontents

\section{Introduction}

The study of the magnetic polarizability of the neutron is an area of ongoing experimental and theoretical interest. Measurement of this quantity remains challenging with considerable uncertainties~\cite{Kossert:2002jc,Kossert:2002ws,Griesshammer:2012we}, although improvement has been seen in recent years~\cite{Myers:2014ace}. There is scope for lattice QCD to make important predictions in this area.
\par
The approach used here to calculate this quantity on the lattice is the uniform background-field method~\cite{Smit:1986fn,Martinelli:1982cb,Burkardt:1996vb}. A $U(1)$ phase factor on the gauge links induces an external magnetic field across the entirety of the lattice. The external field causes an energy shift from which the magnetic polarizability can be determined by use of the energy-field relation~\cite{Martinelli:1982cb,Bernard:1982yu,Burkardt:1996vb,Tiburzi:2012ks,Primer:2013pva,Chang:2015qxa}
\begin{align}
E(B) = m + \vec{\mu}\vdot\vec{B} + \frac{\abs{qe\,B}}{2\,m}-\frac{4\,\pi}{2}\,\beta\,B^2 + \order{B^3},
\label{eqn:n:EB}
\end{align} 
where $m$ is the mass, and $\vec{\mu}$ and $\beta$ are the magnetic
moment and magnetic polarizability respectively. Note that the
$\abs{qe\,B}/2m$ term corresponds to the lowest Landau energy and is
only present for charged hadrons. There is in principle a tower of
Landau levels,  $(2\,n+1)\abs{qe\,B}/2m$ for
$n=0,1,2,\,\dots$~\cite{QFTZuber}. 

At first glance, the method is simple; we can fit the linear and
quadratic coefficients of the resulting energies as a function of
field strength to extract the magnetic moment and
polarizability~\cite{Martinelli:1982cb,Tiburzi:2012ks}. However, baryon correlation
functions suffer from a rapidly decaying signal-to-noise problem \cite{Parisi:1983ae}. This
makes the extraction of the magnetic polarizability using standard nucleon
interpolating fields challenging as it appears at second order in the energy expansion, as demonstrated by previous studies~\cite{Burkardt:1996vb,Tiburzi:2012ks,Primer:2013pva,Chang:2015qxa}.

The application of three-dimensional gauge-covariant Gaussian smearing on the quark fields at the source and/or
sink is highly effective at isolating the nucleon ground state in pure
QCD calculations. However, the presence of a uniform magnetic field
alters the physics, breaking three-dimensional spatial symmetry and
introducing electromagnetic perturbations into the dynamics of the
charged quarks.

Under a uniform magnetic field, in the absence of QCD interactions,
each quark will have a Landau energy proportional to its charge. When
QCD interactions are enabled, the quarks will hadronize, such that (in
the confining phase) the Landau energy corresponds to that of the
composite particle. In particular, as the neutron has zero charge, the $ddu$
quarks must combine such that the overall Landau energy vanishes.

It is clear that the QCD and magnetic interactions compete with each
other in the confining phase. Indeed, there is evidence that residual
Landau mode effects remain when the strong interaction is turned on
\cite{Primer:2013pva,Bruckmann:2017pft}. Here, we explore the idea of
using quark operators on the lattice that capture both of these
forces, choosing asymmetric source and sink operators to provide
better overlap with the energy eigenstates of the neutron in a
background magnetic field.

At the source, we capture the QCD dynamics by using spatial
Gaussian smearing, tuned to maximise overlap with the nucleon ground
state at zero magnetic-field strength. At the sink, we seek to encode
the physics associated with the magnetic field by using a projection
operator constructed from the eigenmodes associated
  with the quark lattice Landau levels.  As discussed in
  Sec.~\ref{sec-LL}, these Landau modes correspond to the

eigenmodes of the two-dimensional $U(1)$ lattice
Laplacian~\cite{Kamleh:2017yjx}. Calculations are performed at
multiple quark masses in order to enable a chiral extrapolation to the
physical regime.

\section{Background Field Method}
To simulate a constant magnetic field along a single axis, the background-field method is used~\cite{Smit:1986fn}. To derive this technique on the lattice, consider the continuum case. In the continuum, a minimal electromagnetic coupling is added to form the covariant derivative
\begin{align}
	D_\mu = \partial_\mu + iqe\,A_\mu.
\end{align}
Here, $A_\mu$ is the electromagnetic four potential, and $qe$ is the charge
on the fermion field. The equivalent modification on the lattice is to
multiply the usual gauge links by an exponential phase factor 
\begin{align}
	U_\mu(x) \rightarrow U_\mu(x)\,e^{iaqeA_\mu(x)}.
\end{align}
A uniform magnetic field along the $\hat{z}$ axis is obtained (in the continuum) using
\begin{subequations}
\begin{align}
	\vec{B} &= \curl{\vec{A}}  \\
	B_z &= \partial_x\,A_y - \partial_y\,A_x,
\end{align}
\end{subequations}
which does not uniquely specify the electromagnetic potential. The choice made over the interior of the lattice is $A_x = -B\,y$. This gives a constant magnetic field of magnitude $B$ in the $+\hat{z}$ direction. In order to maintain the constant magnetic field across the edges of the lattice where periodic boundary conditions are in effect, we set $A_y = +B N_y x$ along the boundary in the $\hat{y}$ dimension. This then induces a quantization condition for the uniform magnetic-field strength~\cite{Primer:2013pva}
\begin{align}
	qe\,B\,a^2 = \frac{2\,\pi\,k}{N_x\,N_y}.
	\label{eqn:qc}
\end{align}
Here, $a$ is the lattice spacing, $N_x$ and $N_y$ are the spatial dimensions of the lattice, and $k$ is an integer specifying the field quanta in terms of the minimum field strength.
\par
In this work, the field quanta $k$ is set in units of the charge of the down quark, i.e. $q=-1/3$. Hence, a field with $k_d=1$ will be in the $-\hat{z}$ direction and aligned with spin-down components.

\section{Quark Operators}

We explore the use of asymmetric source and sink operators in order to construct zero-momentum projected correlation functions which have greater overlap with the energy eigenstates of the neutron in a background magnetic field. This allows us to emulate the dominant QCD and magnetic effects separately.
\par
We consider fixed boundary conditions in the time direction and place the source at $N_t/4 = 16$.

\subsection{Gaussian smeared source}
\label{sec:source-selection}
\begin{figure}[t]
	\centering
        \includegraphics[width=\columnwidth]{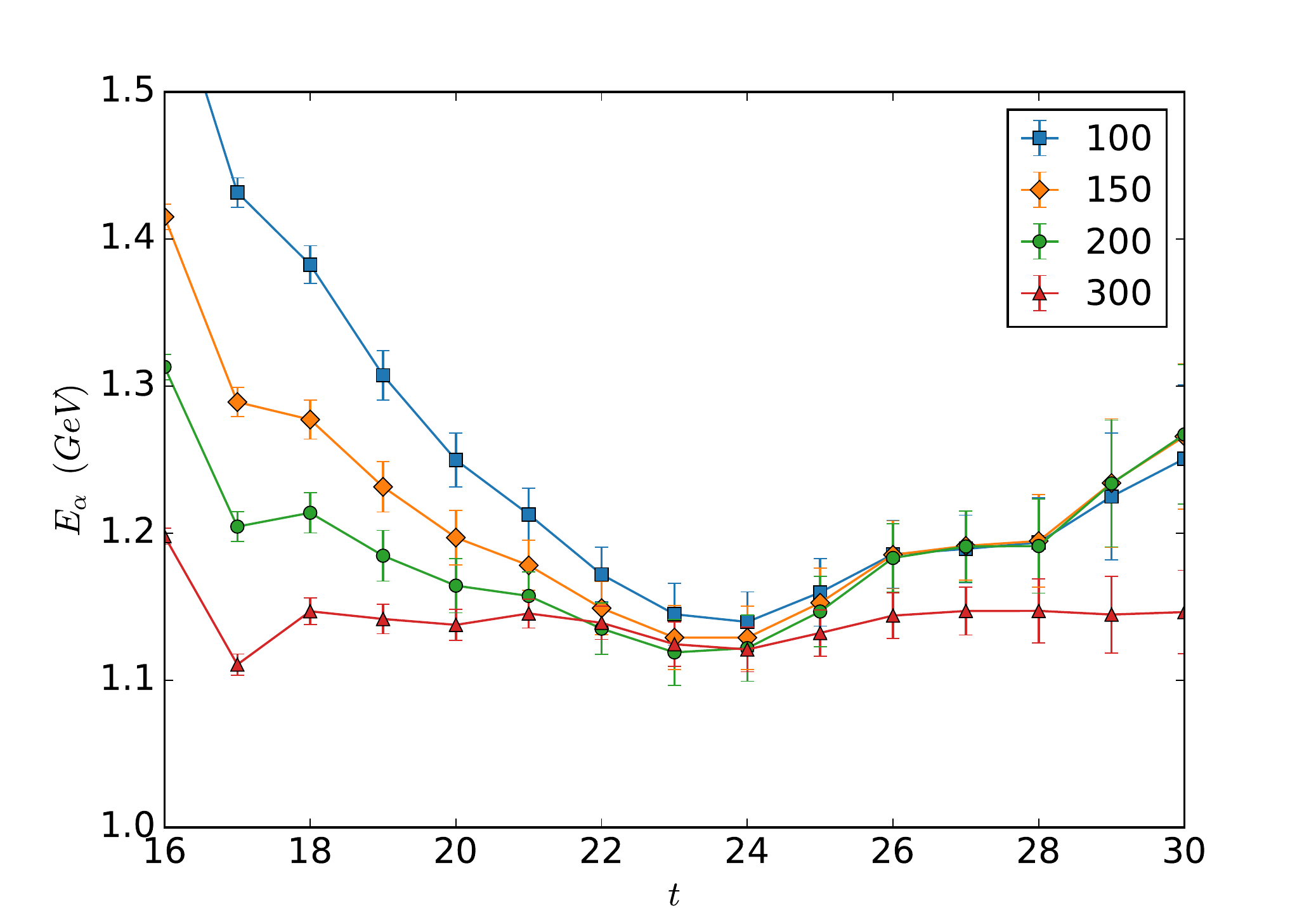}        
	\caption{Neutron zero-field effective mass from smeared source to point sink correlators for various levels of covariant Gaussian smearing at the source. The source is at $t=16$.}
	\label{fig:BF0-manysource}
\end{figure}
\par

A smeared source is used to provide a representation of the QCD
interactions, while the sink is used to capture the physics associated
with the magnetic field.  Several levels of source smearing are
investigated at \hbox{$B=0$} in order to isolate the QCD nucleon ground
state. For \hbox{$m_\pi = 411$ MeV}, $300$ sweeps of standard
Gaussian smearing is optimal as illustrated in \Fig{fig:BF0-manysource}. An identical process is followed at each of
the quark masses producing optimal smearings of $N_{sm} = 150, 175, 300, 350$ for masses $m_\pi=702,570,411,296$ MeV respectively.

\subsection{Landau mode quark sink}
\label{sec-LL}

In the absence of QCD interactions, the charged quarks will each have
an associated Landau level. To capture the physics associated with the
uniform background magnetic field, we apply a quark level $U(1)$
Landau mode projection at the sink. To motivate this sink projection,
we briefly review the relevant Landau mode physics and the relation to the continuum
Dirac equation.
A charged scalar particle which sits in a uniform magnetic field will
have an associated Landau energy which is proportional to its
charge. In the non-relativistic approximation, the energy spectrum of
a charged particle in a constant magnetic field along the $\hat{z}$
direction is equivalent to that of a harmonic oscillator, $E_n = (n +
\frac{1}{2})\, \omega,$ where $\omega = |qeB|/m$ is the classical
cyclotron frequency. In the infinite-volume limit,
each energy level is infinitely degenerate. 

The relativistic generalization of the Landau energy levels for a
fermion commences with the Dirac operator coupled to electromagnetism
\begin{align}
  \slashed{D} = \gmu D_\mu = \gmu \left( \pmu + iqe\,\Amu \right) \, .
\end{align}
The second-order equation for a Dirac spinor $\psi$ is
\begin{align}
  \left( D^2 + \frac{1}{2}qe\,\smunu\Fmunu + m^2 \right)\, \psi = 0 \, ,
\end{align}
such that for a constant background magnetic field $\vB$ (and in a suitable spinorial representation) we have
\begin{align}
  \left( D^2 + qe
  \begin{bmatrix}
    \vs \cdot \vB & 0 \\
    0 & \vs \cdot \vB
  \end{bmatrix}
  + m^2 \right)\,\psi = 0 \, .
\end{align}
Choosing $\vB = B\zh$ in the $\zh$ direction and introducing a
spin-polarization factor, $\alpha = \pm 1$, the equation for each
spinor component $\psi_\tau$ is
\begin{align}
  \left( D^2 + \alpha\,qeB + m^2 \right)\,\psi_\tau = 0 \, ,
\end{align}
with $\alpha = (-1)^{(\tau-1)}$.
The eigenenergies as a function of the mass, $m$; field strength, $B$;
spin polarization, $\alpha$; and momentum in the $z$ direction, $p_z$, are given by
\cite{QFTZuber}
\begin{align}
E^2(B) = m^2 + \abs{qe\,B}\,\left(2\,n + 1 - \alpha\right) + p_z^2,
\end{align}
with $n$ describing the quantized energy level, that is, the relativistic Landau energy. The key point here is that, while the eigenenergies depend on the spin-coupling term, the basis of eigenmodes of
the operator $\left( D^2 + \alpha\,qeB + m^2 \right)$ is independent of the
constant terms $\alpha\,qeB$ and $m^2$, depending only on the covariant
Laplacian $D^2 = D^\mu D_\mu$.  
Hence, on a discrete lattice, the Landau modes for a charged Dirac particle with $\vB= B\, \zh$ correspond to the eigenmodes of the two-dimensional $U(1)$ gauge-covariant lattice
Laplacian
\begin{align}
 \Delta_{\vec{x},\vec{x}\!\:{}'} = 4\delta_{\vec{x},\vec{x}\!\:{}'}
 \!-\!\!\!
 \sum_{\mu=1,2}\!U^{B}_\mu(\vec{x})\,\delta_{\vec{x}+\hat{\mu},\vec{x}\!\:{}'}
 +
 U^{B\dag}_\mu(\vec{x}\!-\!\hat{\mu})\,\delta_{\vec{x}-\hat{\mu},\vec{x}\!\:{}'}
 \, , \label{eq:lap2d}
\end{align}
where $U^{B}_\mu(\vec{x})$ contains the same $U(1)$ phases as
applied in the full lattice QCD calculation.
On a finite-volume lattice, the degeneracy of the lattice Landau modes is finite
and is dependent on the product $qe\,B$ of the charge and magnetic-field
strength.  This is in contrast to the infinite
degeneracy of the infinite volume. In particular, the
lowest Landau level on the lattice has a degeneracy equal to the
magnetic flux quanta $\abs{k}$ defined in Eq.~(\ref{eqn:qc}).

The lowest Landau mode in the continuum takes a Gaussian form,
$\psi_{\vec{B}}(x,y) \sim e^{-|qeB|\,(x^2+y^2)/4}.$ It has been noted
elsewhere~\cite{Tiburzi:2012ks,Bignell:2017lnd} that in a finite
volume the periodicity of the lattice causes the wave function's form
to be altered. We can calculate the eigenmodes of the 2D Laplacian in
\eqnr{eq:lap2d} and project at the quark level. Define a projection
operator onto the lowest $n$ eigenmodes $|\,\psi_{i, \vec{B}}\rangle$
of the two-dimensional (2D) Laplacian as 
\begin{align}
P_n = \sum_{i=1}^{n}\, |\,\psi_{i, \vec{B}}\rangle\,\langle\psi_{i, \vec{B}} \,|\,. 
\end{align}
A coordinate-space representation of this two-dimensional projection
operator is applied at the sink to the quark propagator
\begin{align}
 S_n(\vec{x},t;\vec{0},0) = \sum_{\vec{x}\!\:{}'} P_n(\vec{x},\vec{x}\!\;{}')\,S(\vec{x}\!\;{}',t;\vec{0},0)\,, \label{eq:landausink}
\end{align}
where $n = \abs{3\,q_f\,k_d}$ modes for the lowest Landau level.
\par
The $U(1)$ Laplacian is not QCD gauge covariant, and hence we fix the
gluon field to Landau gauge and apply the appropriate gauge rotation
to the quark propagator before projecting. However, as the hadronic
correlation function (and ground state energy) is gauge invariant,
using a gauge-fixed sink operator can only effect the overlap with the
ground state, which has the potential to improve the final precision
of our result.

\subsection{One-dimensional spatial modulation}

\begin{figure}[t]
	\centering
	\includegraphics[width=\columnwidth]{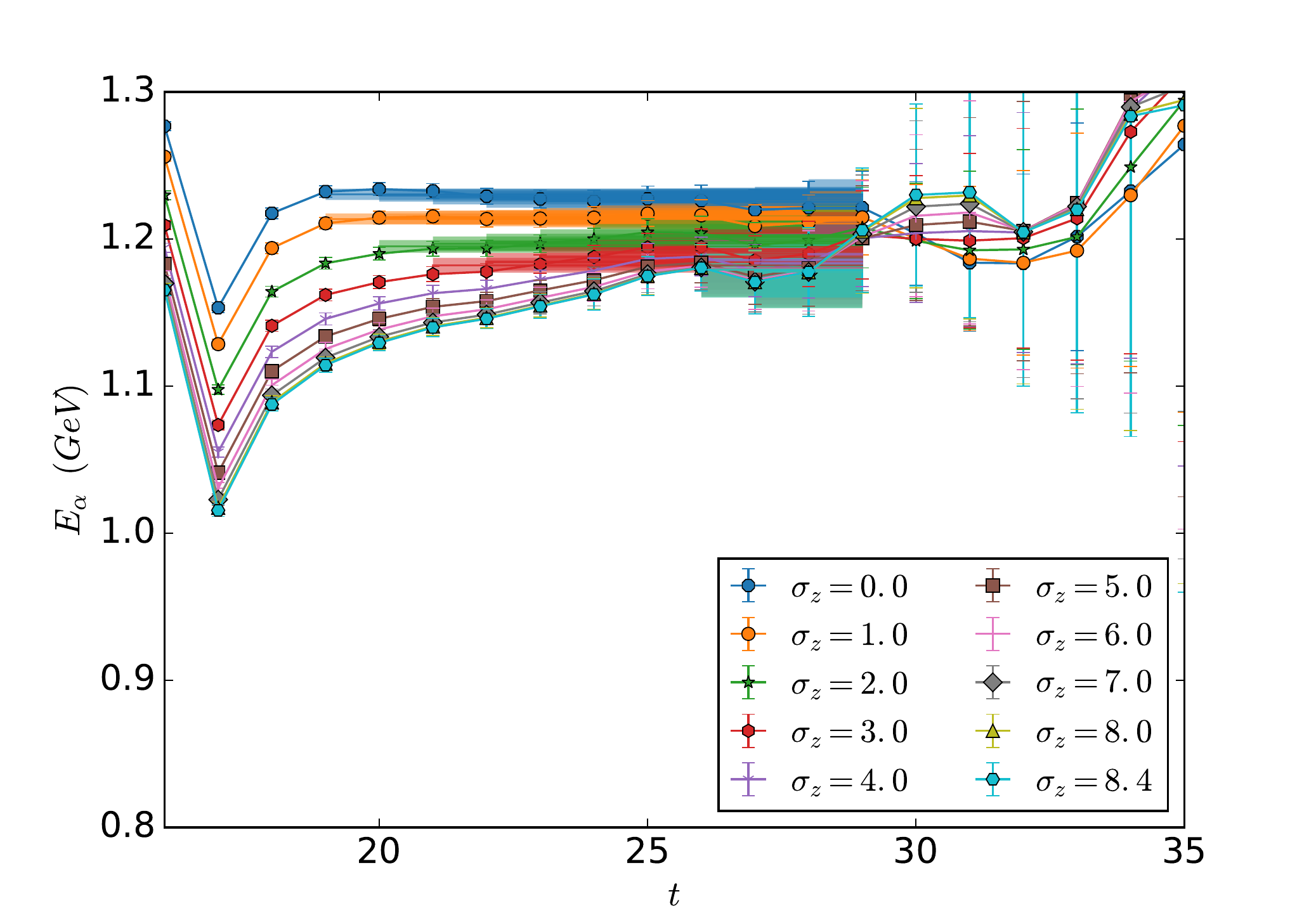}
	\caption{Anti-aligned effective energy of the neutron in the largest field strength, $\abs{k}=3$, for $U(1)$ Landau-projected sinks at $m_\pi = 411$ MeV. Consecutive fits ending at $t=29$ where all effective masses agree with $\chi^2_{dof} \le 1.2$ are shown.}
	\label{fig:TBMSE-BF3-AA-29}
\end{figure}
\begin{figure}[t]
	\centering
	\includegraphics[width=\columnwidth]{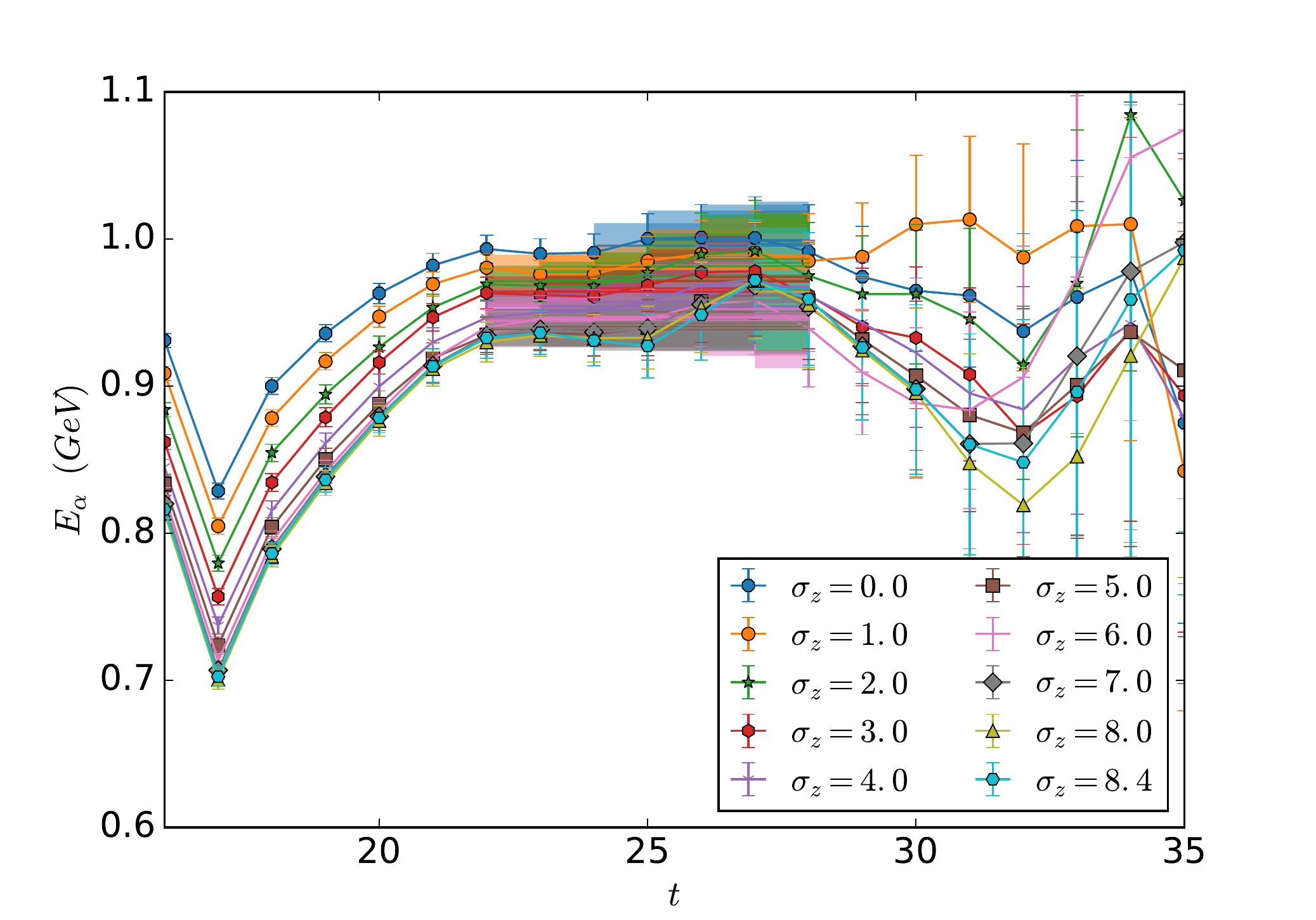}
	\caption{Aligned effective energy of the neutron in the smallest field strength, $\abs{k}=1$, for $U(1)$ Landau-projected sinks at $m_\pi = 296$ MeV. Consecutive fits ending at $t=28$ where all effective masses agree with $\chi^2_{dof} \le 1.2$ are shown.}
	\label{fig:TBMSE-BF3-A-28_k4}
\end{figure}
The eigenmodes of the two-dimensional $U(1)$ Laplacian have no dependence on the $z$ coordinate. Using this freedom, we can apply a functional form to vary the spatial extent of the $U(1)$ Landau projection in the $\hat{z}$ direction, an idea analogous to standard Gaussian smearing. We modulate the $z$ dependence of the projected quark propagator with a normalized Gaussian
\begin{equation}
\phi_{\sigma}(z) = \frac{1}{\sigma\sqrt{2\pi}}\exp(-\frac{z^2}{2\sigma^2}),
\label{eqn:phisigmaz}
\end{equation}
where the width parameter $\sigma\equiv\sigma_z$ controls the spatial extent in the $z$ direction. After the $U(1)$ Landau mode projection has been applied at the sink to the quark propagator as in \eqnr{eq:landausink}, the gauge-fixed propagator is then averaged over the $z$ dimension using the modulation function as a weighting,
\begin{align}
S_{n,\sigma}(x,y,z,t;\vec{0},0) = \sum_{z'} \phi_\sigma(z-z')\,S_n(x,y,z',t;\vec{0},0)\,. \label{eq:sigmasink}
\end{align}
We define the special case $\sigma_z = 0$ to indicate that no $z$ modulation is applied, which is equivalent to choosing $\phi_{\sigma=0}(z) = \delta(z'-z),$ such that $S_{n,0} \equiv S_n.$

Different spatial extents change the coupling to each of the energy eigenstates. The lowest lying level is dominant in the long Euclidean time limit. To determine which spatial extent provides the greatest overlap with the lowest lying energy level, many choices of $\sigma_z$ are investigated simultaneously~\cite{Mahbub:2010rm}.

The magnetic-field orientation and neutron spin polarization can be
chosen independently to be in the positive or negative $z$
direction. In order to efficiently extract the magnetic
polarizability, combinations of correlation functions with differing
magnetic-field orientation and spin-polarization alignments are used
to create spin and magnetic field aligned and anti-aligned correlation
functions. These are the energies which will be examined in order to
optimize the quark sink. 

The quark sink selected is the one which has the longest plateau when
fitting backward in Euclidean time from where all of the correlators
agree.  In evaluating this extent, the $\chi^2_{dof}$ is determined
via a consideration of the full matrix of covariances between
different time slices under consideration and we employ an upper limit
of 1.2.  The sink-projected correlator that has converged the earliest
is considered optimal.  This process is undertaken for each
combination of field strength and aligned or anti-aligned energies.
\Fig{fig:TBMSE-BF3-AA-29} shows an example of this process for
the $m_\pi=411$ MeV neutron and the largest magnetic field considered
with $\abs{k}=3$. It is quite clear that all the sink projections
agree by $t=29$ and that $\sigma_z=0,1$ both produce excellent
early plateaus. 
\Fig{fig:TBMSE-BF3-A-28_k4} shows the aligned
energies for $m_\pi = 296$ MeV in the smallest field strength. In this
case, there is no clear longest plateau.
In cases like this where multiple $\sigma_z$ sink projections are
allowed by both length and the $\chi^2_{dof}$, the full process for
calculating the magnetic polarizability is performed for each value of
$\sigma_z$.  The resulting magnetic polarizability values are averaged
to give a combined statistical error as well as a systematic error
associated with the range of allowed $\sigma_z$. 

\begin{figure}[t]
	\centering
	\includegraphics[width=\columnwidth]{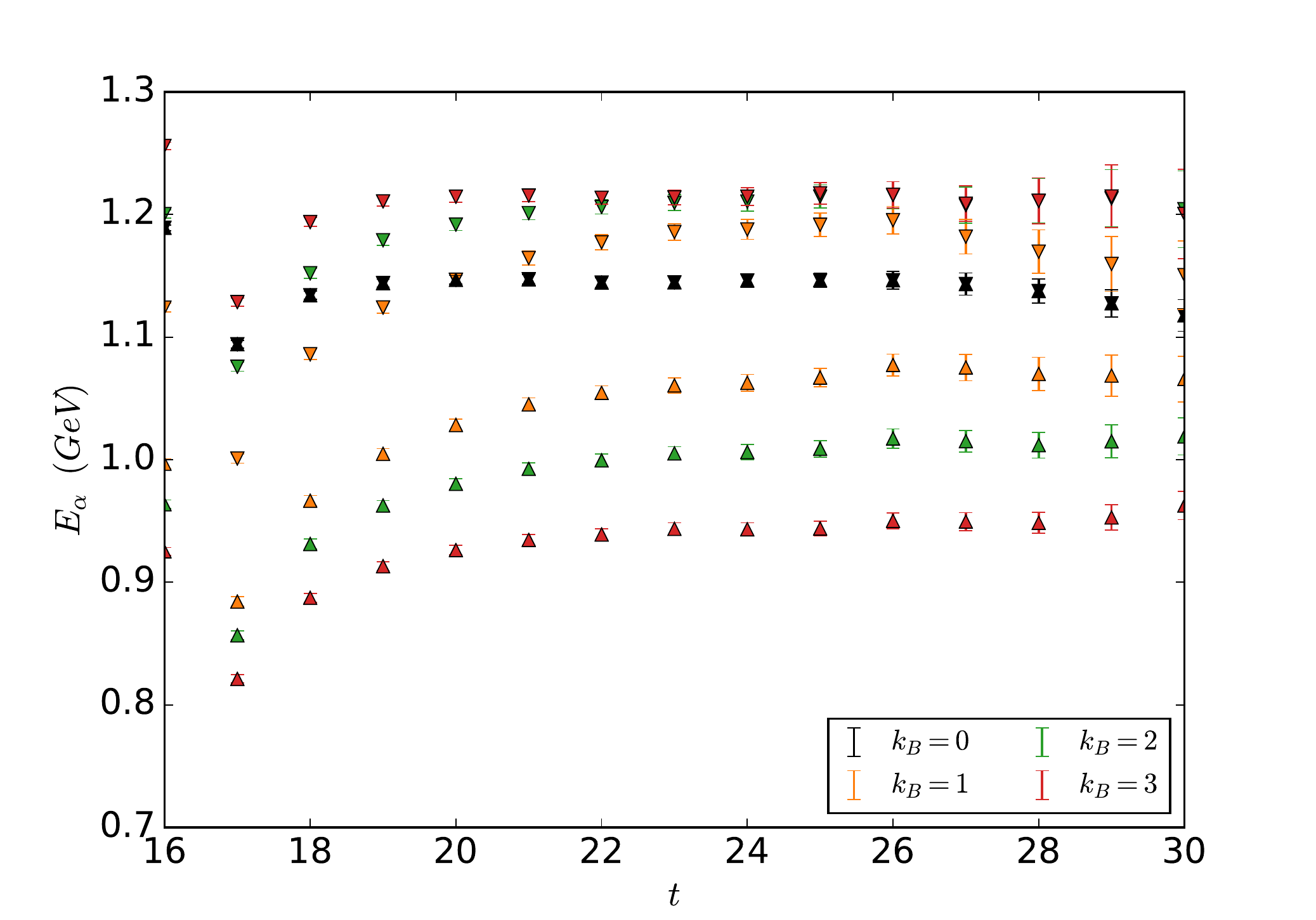}
	\caption{Aligned (up arrows) and anti-aligned (down arrows) effective energies of the \hbox{$m_\pi = 411$ MeV} neutron using a $U(1)$, $\sigma_z=1.0$ Landau mode sink projection. Three non-zero field strength energies and the zero-field mass are illustrated.}
	\label{fig:E(B)}
\end{figure}

In general, small $\sigma_z$ values, $\sigma_z = 0,1,2$, are preferred
across multiple pion masses, field strengths and aligned or
anti-aligned combinations.  These sink projections provide a good
representation of the neutron ground state in a background magnetic
field as can be seen by the plateau behaviour in the energy of the
neutron in \Fig{fig:E(B)}.

\color{black}

This result represents a significant advance in the determination of magnetic polarizabilities. For the first time clear plateaus are identified, a direct result of our consideration of Landau modes at the quark level.

\begin{figure}[t]
	\centering
        \includegraphics[width=\columnwidth]{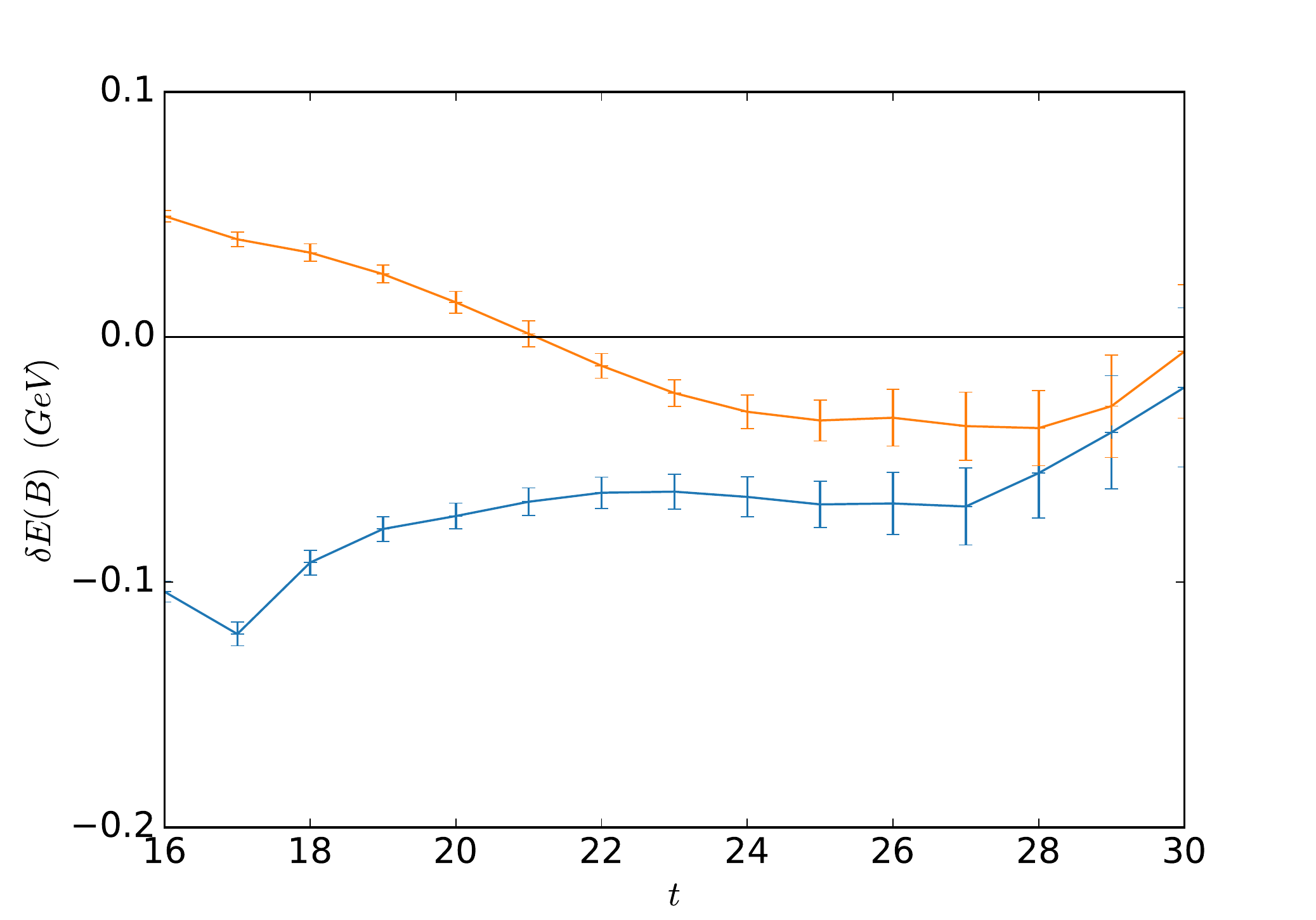}
	\caption{The magnetic polarizability effective energy shift at the largest field strength for the neutron as a function of Euclidean time (in lattice units), using a smeared source. A point sink (orange) and a $U(1)$ Landau mode quark sink (blue) are illustrated.}
	\label{fig:dEcomp}
\end{figure}

\section{Magnetic Polarizability}
\label{sec:MagPol}
\subsection{Formalism}
Recalling the energy-field relation of \eqnr{eqn:n:EB}, we note that a combination of energies at different spin orientations and field strengths can be used to isolate the neutron magnetic polarizability $\beta,$ 
\begin{align}
	\delta E(B) &= \frac{1}{2}\,\left[ \left( E_\uparrow(B) - E_\uparrow(0) \right) + \left( E_\downarrow(B) - E_\downarrow(0) \right) \right] \nonumber \\
	&= - \frac{4\,\pi}{2}\,\beta\,B^2 + \order{B^4},
	\label{eqn:dE}
\end{align}
noting that as $q = 0$ for the neutron the Landau energy term vanishes. Here, the arrows denote the neutron spin polarization along the $\hat{z}$ axis.

This method of isolating the polarizability term is valid, but in practice due to the cancellation of correlated fluctuations on a common ensemble of lattice configurations it is much more effective to take ratios of appropriate spin-up ($+s$) and spin-down ($-s$) correlators. We can also average over both positive ($+B$) and negative ($-B$) magnetic-field orientations to provide an improved unbiased estimator. Thus, we define the spin-field aligned correlator by
\begin{align}
	G_{\upup}(B) = G(+s,+B) + G(-s,-B),
\end{align}
and the spin-field anti-aligned correlator by
\begin{align}
G_{\updown}(B) = G(+s,-B) + G(-s,+B).
\end{align}
The spin-field aligned and anti-aligned correlators, combined with the spin-averaged zero-field correlator are used to form the ratio
\begin{align}
	R(B,t) = \frac{ G_{\upup}(B,t)\,G_{\updown}(B,t) }{G(0,t)^2}
        \, .
	\label{eqn:RBt}
\end{align}
The product of the spin-field aligned and anti-aligned correlators
yields an exponent that is the sum of the respective energies $\sim E_{\upup}+E_{\updown},$ removing the contribution from the magnetic moment term. Our calculation is systematically 
improved by including the contributions from all four field and spin pairings, such that
upon taking the effective energy we obtain the desired energy shift,
\begin{align}
	\delta E(B,t) &= \frac{1}{2}\,\frac{1}{\delta t}\log( \frac{R(B,t)}{R(B,t+\delta t)} )\nonumber
		\label{eqn:RdE} \\
	&= - \frac{4\,\pi}{2}\,\beta\,B^2 + \order{B^4}.
\end{align}
Note that we define the magnetic field $\pm B$ to be that experienced by the nucleon, and is hence related to the down quark magnetic field by a factor of $-3.$

Any correlated QCD fluctuations between the finite
  field strength and zero-field effective energies are significantly
  reduced by taking the ratio in Eq.~(\ref{eqn:RBt}).  As the zero-field
  correlator does not have a Landau level, the $U(1)$ eigenmode
  projection technique is not applied, and we use a standard point sink instead. This motivates the source tuning process outlined in Sec.~\ref{sec:source-selection}. By using a source optimized for the
  zero-field neutron in the denominator of Eq.~(\ref{eqn:RBt}), the
  onset of plateau behavior in the effective energies occurs at an
  early Euclidean time.
This improved method is particularly important as the polarizability is at second order in $B,$ and as such at these small field strengths, its contribution to~\eqnr{eqn:n:EB} is small. It is essential to have a precise determination of the polarizability energy shift. The efficiency of the Landau mode sink projection can be seen in \Fig{fig:dEcomp} where the energy shift for a standard, point sink is compared to a $U(1)$ Landau mode sink projection; the latter is seen to display better plateau behaviour.

\begin{figure}[t]
	\centering
	\includegraphics[width=\columnwidth]{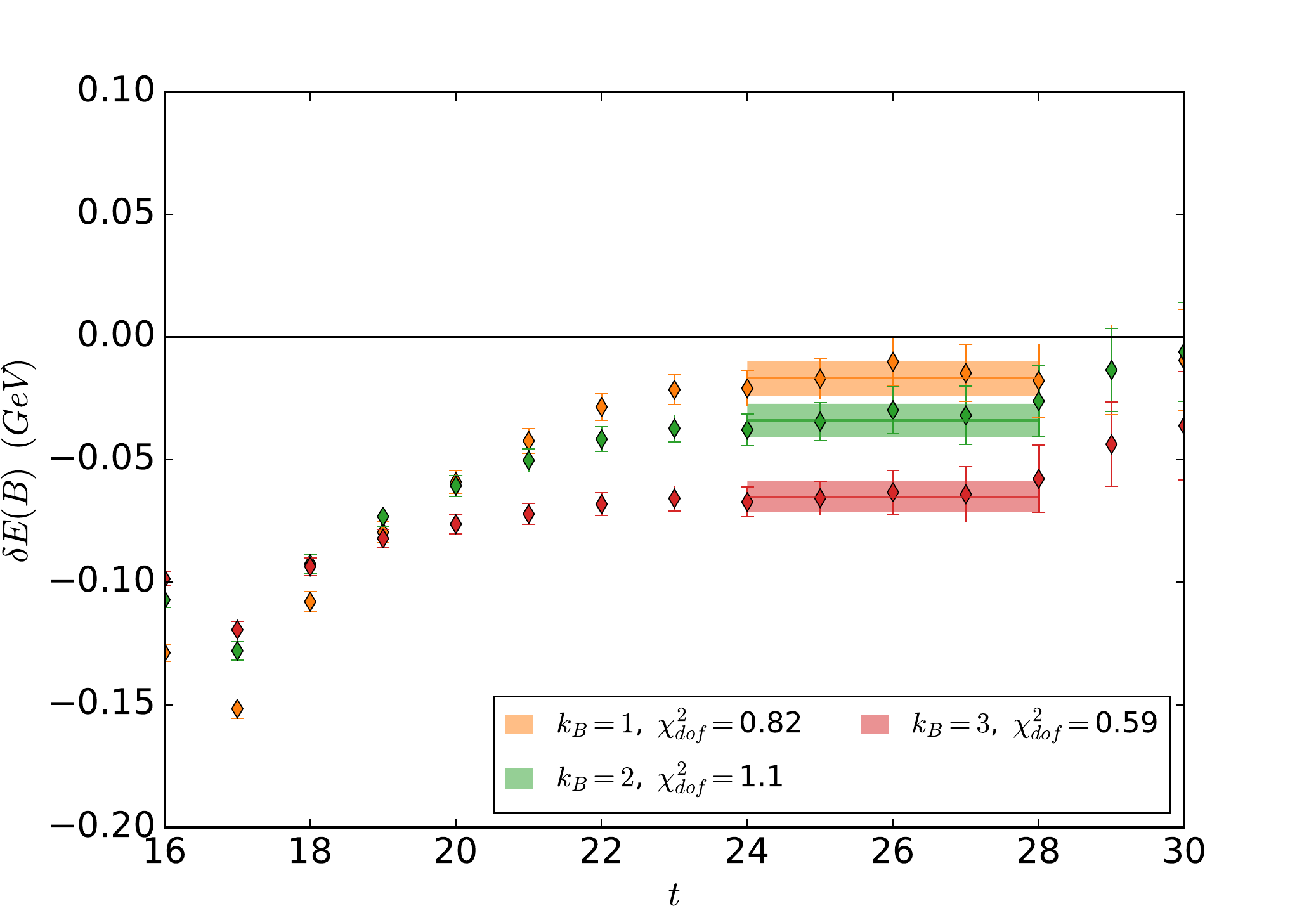}
	\caption{The magnetic-polarizability effective-energy shift for the $m_\pi = 411$ MeV neutron as a function of Euclidean time (in lattice units), using a smeared source and $\sigma_z=1.0$ $U(1)$ Landau mode sink projection. Results for field strengths $k_B=1,2,3$ are shown, with the magnetic-field strength increasing away from zero. The selected fits and $\chi^2_{dof}$ are also illustrated.}
	\label{fig:dEplat-22-25-sz1}
\end{figure}
\begin{figure}[t]
	\centering
	\includegraphics[width=\columnwidth]{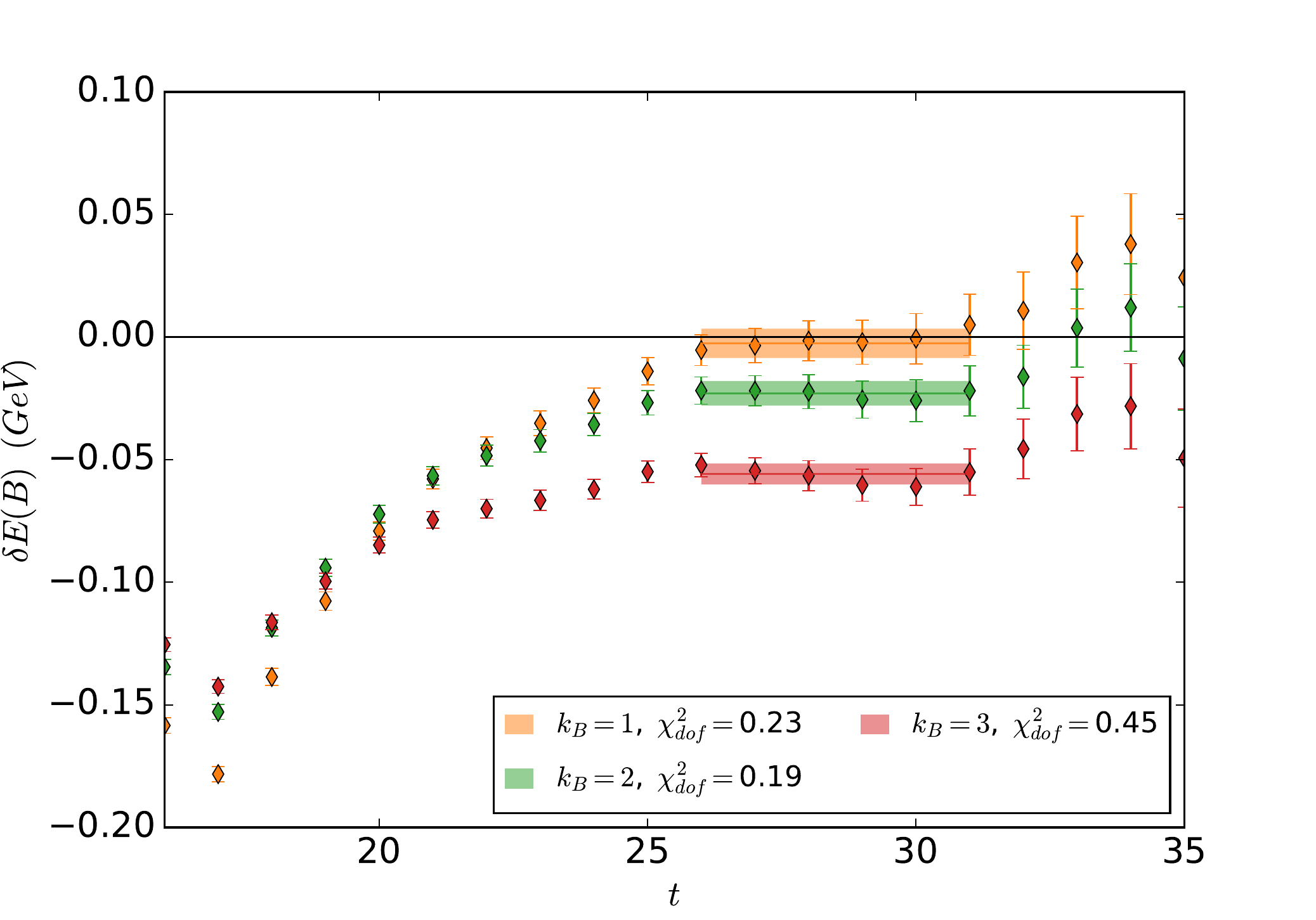}
	\caption{The magnetic-polarizability effective-energy shift for the $m_\pi = 702$ MeV neutron as a function of Euclidean time (in lattice units), using a smeared source and $\sigma_z = 0.0$ $U(1)$ Landau mode sink projections. Symbols are as described in \Fig{fig:dEplat-22-25-sz1}.}
	\label{fig:dEplat-22-25-sz1b}
\end{figure}
\par

\subsection{Simulation Details}

In this work, $2+1$ flavour dynamical gauge configurations provided by the PACS-CS~\cite{Aoki:2008sm} group through the ILDG~\cite{Beckett:2009cb} are used. These have a clover fermion action and Iwasaki gauge action with a physical lattice spacing of $a = 0.0907(13)$. Four values of the light quark hopping parameter $k_{ud} = 0.13700,\,0.13727,\,0.13754,\,0.13770$ are considered, corresponding to pion masses of \hbox{$m_\pi = 702,\,570,411,\,296$} MeV respectively. The lattice spacing for each mass was set using the Sommer scale with $r_0 = 0.49$ fm. The lattice volume is $L^3 \times T =32^3 \times 64$ and the ensemble sizes are $399,\,400,\,449,\,400$ configurations respectively.
Source locations were systematically varied in order to produce large distances between adjacent source locations. Starting from an initial source location at $\vectwo{0}{16}$, shifts of $\vectwo{0}{16}$ were applied three times for a total of four source locations. A further set of four shifts starting at $\vectwo{16}{8}$, where only the time component increases by $16$, were also applied. As such, a total of eight sources were used for each configuration.
\par
Correlation functions at four distinct magnetic-field strengths are calculated. To do this, propagators at ten non-zero field strengths, $e\,B = \pm 0.087$, $\pm 0.174$, $\pm 0.261$, $\pm 0.348$, $\pm 0.522$ GeV$^2$, are calculated. These correspond to $k_d = \pm 1, \, \pm 2,\, \pm 3,\,\pm 4,\,\pm 6$ in \eqnr{eqn:qc}. The zero-momentum projected correlation functions contain spin-up and spin-down components.

We note that at the higher field strengths considered here one might be concerned about the validity of the
energy-field expansion of \eqnr{eqn:n:EB}. We can relate the energy-field expansion in \eqnr{eqn:n:EB} to the relativistic energy of a baryon in an external background magnetic
field by considering $E^2(B) - m^2 = (E(B)-m)(E(B)+m)$ and applying the non-relativistic approximation $(E+m)\simeq 2m.$
Thus, $2m/(E+m) \simeq 1$ is a measure of the importance of
relativistic effects.  We find $2m/(E+m)$ to be typically within a few
percent of one for all but the largest field strength.  At the lightest
quark mass the effect can approach 10\%.  However, this is a small
effect in the context of the current statistical uncertainties and
other systematic uncertainties discussed in the following.  Still,
it is an important issue to consider as one moves toward the precision
era of magnetic polarizability calculations in lattice QCD.
It is also important to note that these configurations are electro-quenched; the field exists only for the valence quarks of the hadron. To include the background field at configuration generation time is possible~\cite{Fiebig:1988en} but requires a separate Monte Carlo simulation for each field strength and is hence prohibitively expensive. Separate calculations also destroy the advantageous correlations between the field strengths used when constructing the ratio in \eqnr{eqn:RBt}.
An alternative is to use a reweighting procedure on the gauge-field configurations~\cite{Freeman:2014kka} for the different field strengths $B,$ but this is not performed here.

\section{Fitting}

The energy shift at each field strength has the form specified by \eqnr{eqn:dE}, and as such, we fit with a quadratic term,
\begin{align}
	\delta E(B,t)  = \frac{4\,\pi}{2}\,\beta\,B^2 + \order{B^4}.
	\label{eqn:ndeB}
\end{align}
Figures \ref{fig:dEplat-22-25-sz1}and \ref{fig:dEplat-22-25-sz1b} show fits for the neutron energy shift with a smeared source and a $U(1)$ Landau mode sink projection. For the first time, clear plateaus are present in this difficult-to-obtain quantity. It is required that a plateau be present at each of the three non-zero field strengths in order to proceed to the next stage.
\par
 The plateau does not occur until $t=24$; this region is a common starting point across the heavier masses. The primary cause for this late plateau onset time is the zero-field correlator, which has fundamentally different physics. As such its potential excited state behaviour is different than that of the background-field correlators. Plateaus only form once both correlators have decayed to the ground state.
\par
\begin{figure}[t]
	\centering
	\includegraphics[width=0.475\columnwidth]{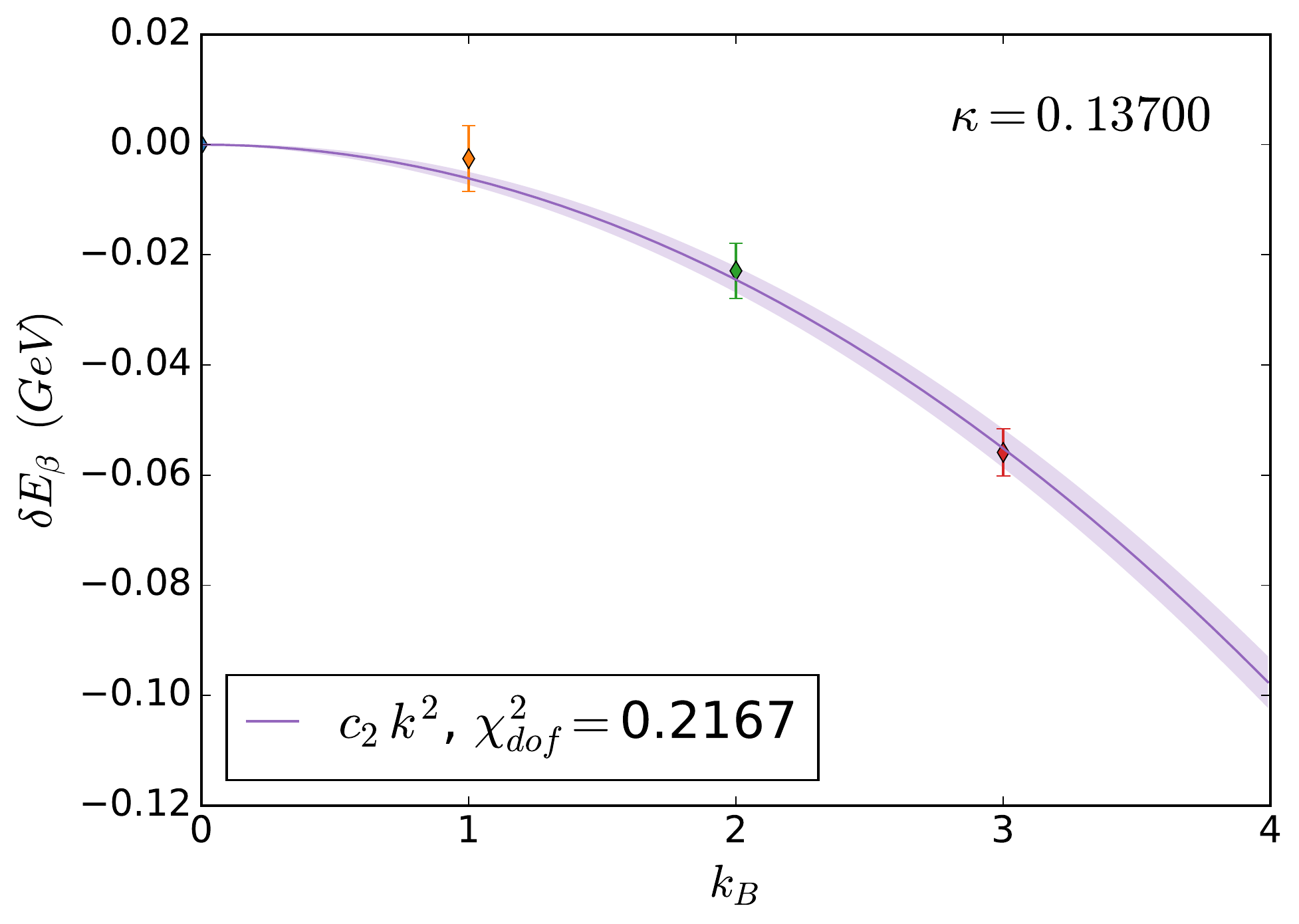}
	\includegraphics[width=0.475\columnwidth]{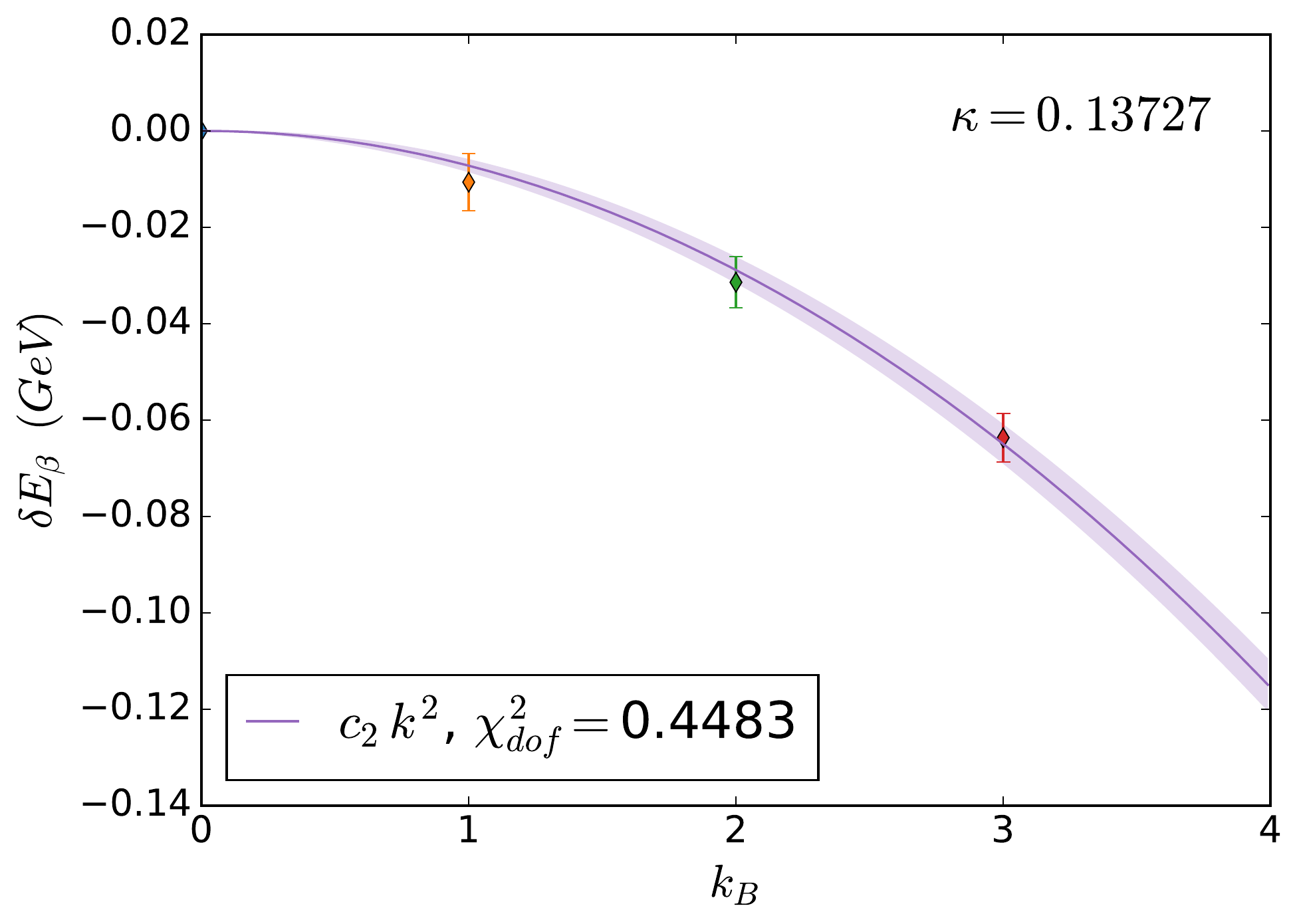}%
        \\
	\includegraphics[width=0.475\columnwidth]{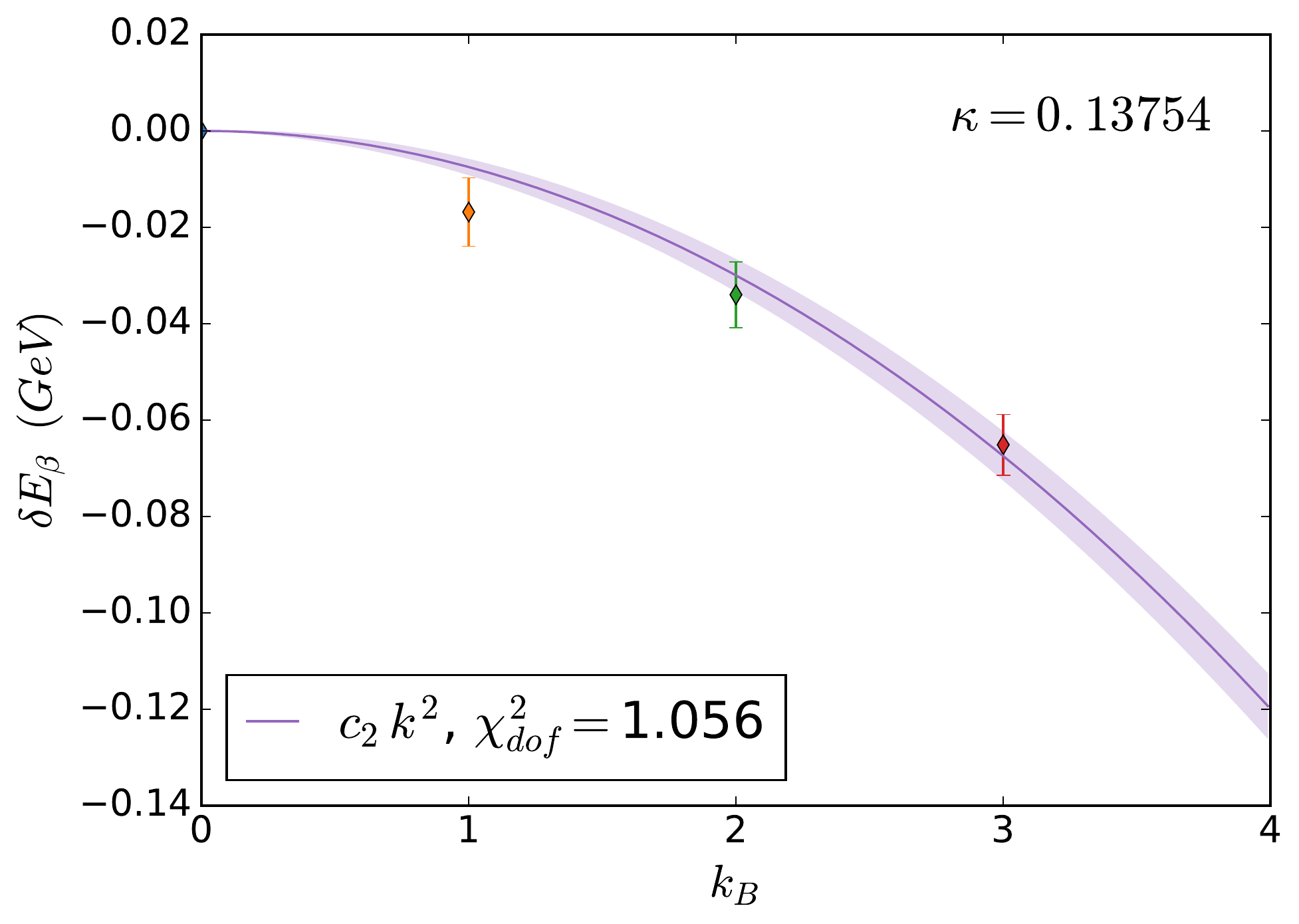}
	\includegraphics[width=0.475\columnwidth]{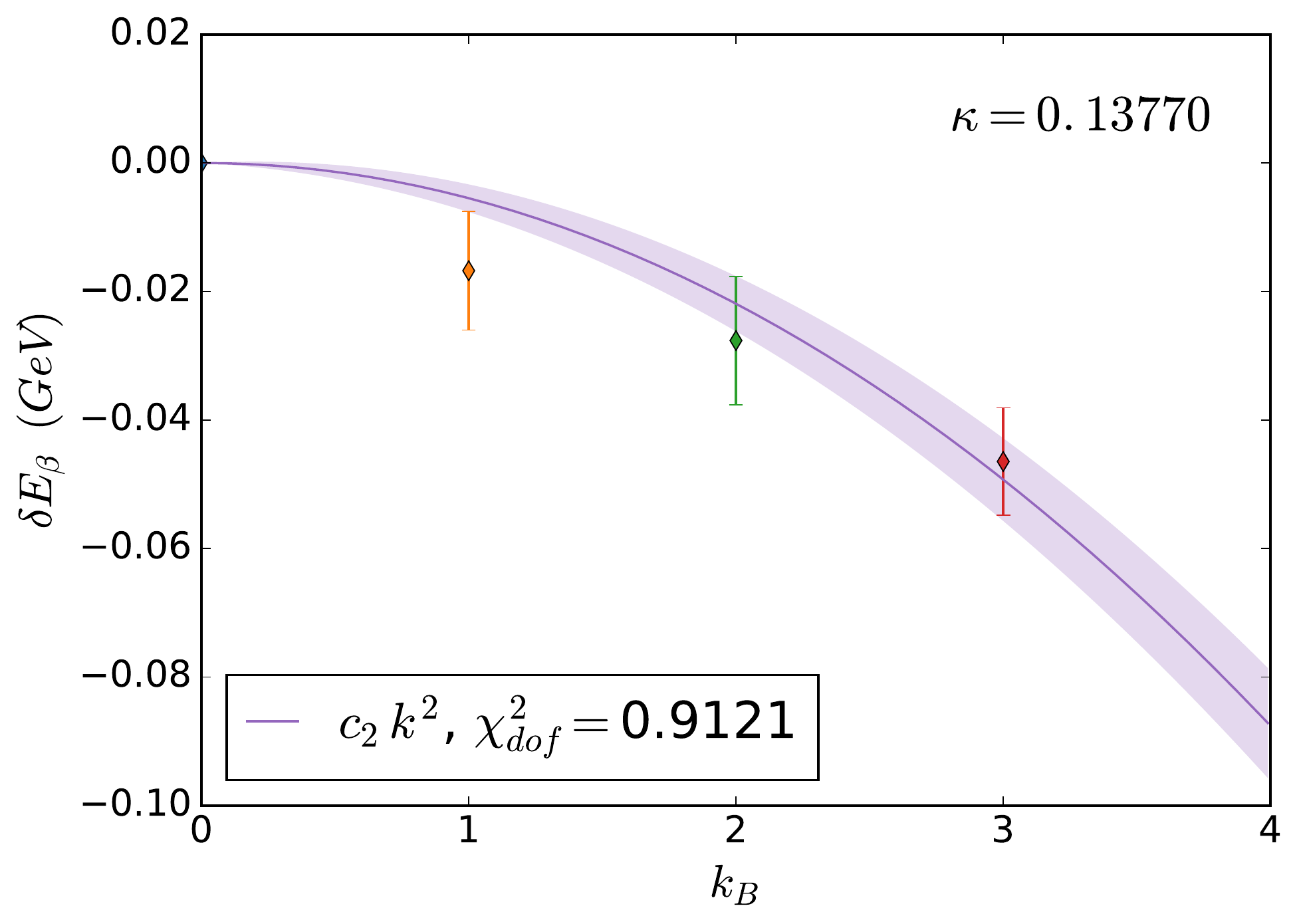}
	\caption{Quadratic fits of the energy shift to the field quanta at each quark mass for the neutron for a single $\sigma_z$ value each.}
	\label{fig:dEBfits}
\end{figure}

The fit performed is as a function of $k_d$, the integer magnetic flux quanta in~\eqnr{eqn:qc},
\begin{align}
  \delta E(k_d) = c_2\,k_d^2.
  \label{eqn:c2}
\end{align}
Here, $c_2$ is the fit parameter, and has units of GeV as these are the units of $\delta E(k_d)$. As a check of the validity of the expansion in \eqnr{eqn:n:EB} and hence the energy shift in \eqnr{eqn:ndeB}, we also perform a quadratic $+$ quartic fit, $c_2\,k_d^2+c_4\,k_d^4,$ where the size of the quartic term provides an estimate of the corrections. It is found that for the two heavier masses, $m_\pi = 702,\, 570$ MeV that the quartic term is indistinguishable from zero, while for $m_\pi = 411$ MeV, the fit is disfavoured by the $\chi^2_{dof}$ of the fit. While the quadratic $+$ quartic fit works for $m_\pi = 293$ MeV, the uncertainties are extremely large, suggesting that such a fit is only possible due to the larger uncertainties associated with lighter quark mass. If \eqnr{eqn:n:EB} were not valid at the field strengths considered herein, a remnant magnetic moment term proportional to $B$ would exist in \eqnr{eqn:ndeB}. It was found that it is possible to fit a purely quadratic term as in \eqnr{eqn:c2} at each pion mass, and the inclusion of a  quartic term is not required, confirming the validity of \eqnr{eqn:n:EB} for the neutron at the field strengths considered in this study. The quadratic fits are displayed in \Fig{fig:dEBfits}.
\par
In order to convert this fit parameter to the physical units of magnetic polarizability, $\text{fm}^3$, \eqnr{eqn:qc} is used to produce the transformation
\begin{align}
\beta = -2 \, c_2\,\alpha\,q_d^2\,a^4\,\left( \frac{N_x\,N_y}{2\,\pi}\right)^2,
\label{eqn:correct-final_beta}
\end{align} 
Here, $\alpha$ is the fine structure constant, $\alpha \approx 1/137$.
\par
The quadratic fitting process uses only energy shifts which have the same spatial modulation of \eqnr{eqn:phisigmaz}. As was seen in \Figtwo{fig:TBMSE-BF3-AA-29}{fig:TBMSE-BF3-A-28_k4}, the spatial extent affects the coupling to the energy eigenstates. It is hence important that when we fit we use the optimal sink projection. This is achieved using a simultaneous investigation of the spatial extent, $\sigma_z$, and field strength. For a specified spatial modulation to be suitable, it must provide early isolation of the eigenstate at each field strength. This isolation is visible in the long plateaus of \Figtwo{fig:TBMSE-BF3-AA-29}{fig:TBMSE-BF3-A-28_k4}.
\par
This is already a strong constraint on the sink choice, but in order to determine where to fit energy shifts for the quadratic fit in $B$, a further constraint is needed. This constraint comes from considering the constant plateau fits to the energy shift at all field strengths. By considering all possible fit windows, we select fit windows where good plateau behaviour exists for all field strengths simultaneously. Good plateau behaviour is characterized by a fit $\chi^2_{dof}$ of less than $1.2$. This process of requiring good plateau behaviour at each field strength simultaneously dramatically reduces the number of possible fit windows. In particular, it is often difficult to obtain acceptable energy shift plateaus for the largest field strength considered.
\par
The final constraint on the fitting process comes from the quadratic fit itself. This fit must also be acceptable having a $\chi^2_{dof} \le 1.2$. If multiple fit windows remain after this process, the one with the longer time extent and $\chi^2_{dof}$'s closest to one are preferred.

\par
Once the specific quadratic fit has been chosen, the magnetic polarizability, $\beta$, is extracted from the quadratic coefficient of the fit. In order to test the presence of higher order terms in the energy shift of \eqnr{eqn:ndeB}, a quartic term is also considered. It is found that the quartic term is not needed in order to fit the energy shifts well with acceptable $\chi^2_{dof}$.

\par
Using the sink eigenmode-projection technique at each quark mass, it is possible to extract magnetic polarizabilities from the fits to the constant energy shift plateaus as a function field strength. Results are presented in Table \ref{tab:polvals2} for the magnetic polarizability of the neutron at each quark mass. Note that for the $m_\pi = 570$ MeV ensemble no systematic error due to the choice of $\sigma_z$ is reported as only $\sigma_z=0$ provided good access to the ground state across all field strengths.

\begin{table}[h!]
	\caption{Magnetic polarizability values for the neutron at each quark mass. Eight sources are used for each quark mass. The numbers in parentheses describe statistical and systematic uncertainties respectively.}
	\label{tab:polvals2}
	\centering
	\begin{ruledtabular}	
		\begin{tabular}{ccccc}
			$\kappa$ & $m_\pi$ (MeV) & $\beta\left( \text{fm}^3 \times 10^{-4}\right)$ & $\chi^2_{dof}$ \\ \hline
			0.13700 & 702 & 1.51(21)(6) & 0.21 \\
			0.13727 & 570 & 1.63(16) & 0.44 \\
			0.13754 & 411 & 1.29(20)(11) & 1.06 \\
			0.13770 & 296 & 1.14(25)(17) & 0.91 \\
		\end{tabular}
	\end{ruledtabular}
\end{table}

\section{Chiral Extrapolation}

\subsection{Formalism}

Chiral effective-field theory ($\chi$EFT) is an important tool for
connecting lattice results to the physical point. The analysis
here follows that of Ref.~\cite{Hall:2013dva} and is summarized
briefly below. 

\begin{figure}[t!]
	\centering
	%\vspace{-30pt}
	\includegraphics[width=\columnwidth,origin=c]{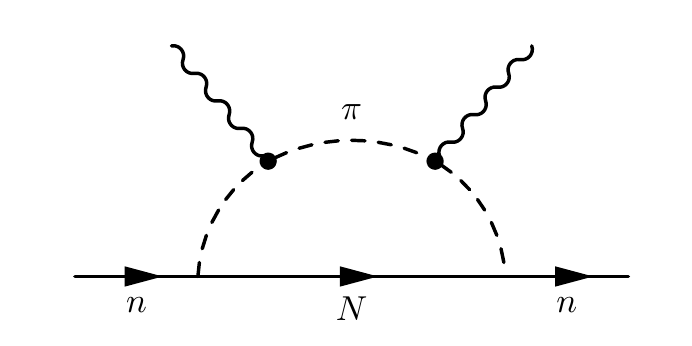}
	%\vspace{0.3cm}
	\caption{The leading-order pion-loop contribution to the magnetic polarizability of the neutron}
	\label{fig:chiEFT_piN}

\end{figure}

% Delta intermediate state
\begin{figure}[t!]
	\centering
	%\vspace{-30pt}
        \includegraphics[width=\columnwidth,origin=c]{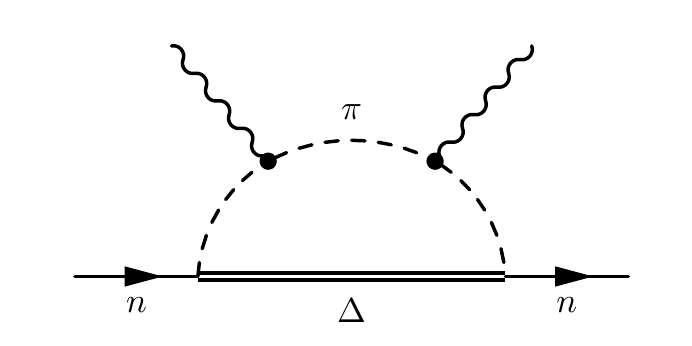}
	%\vspace{0.3cm}
	\caption{Pion-loop contributions to the magnetic
          polarizability of the neutron, allowing transitions to the
          nearby and strongly coupled $\Delta$ baryons.} 
	\label{fig:chiEFT_piDelta}
\end{figure}

We consider the chiral expansion
\begin{align}
\beta \left(m_\pi^2\right) = \beta^{\pi N}\!\left(m_\pi^2\right) 
                           + \beta^{\pi \Delta}\!\left(m_\pi^2\right) + a_0 + a_2\,m_\pi^2 \, .
\label{eqn:chiexpan}
\end{align}
The leading-order loop contributions $\beta^{\pi N}\!\left(m_\pi^2\right)$ and
$\beta^{\pi \Delta}\!\left(m_\pi^2\right)$ are depicted in \Figtwo{fig:chiEFT_piN}{fig:chiEFT_piDelta}.  
The loop integral of \eqnr{eqn:betaPiN} for $\beta^{\pi N}\!\left(m_\pi^2\right)$ contains the
leading non-analytic contribution to the chiral expansion proportional to
$1/m_\pi$~\cite{Lensky:2009uv}.  Similarly, the integral of \eqnr{eqn:betaPiDelta} for $\beta^{\pi
  \Delta}\!\left(m_\pi^2\right)$ accounts for transitions to a Delta baryon.  For a finite
nucleon-Delta mass splitting, $\Delta = M_\Delta - M_N$, this diagram contributes a non-analytic
logarithmic contribution proportional to $(-1/\Delta) \log(m_\pi/\Lambda)$ to
the expansion.  Here, $\Lambda$ is a renormalization scale.
The coefficients $a_0$ and $a_2$ are residual-series coefficients \cite{Leinweber:2003dg} which
will be constrained by our lattice QCD results after they are corrected to infinite volume.  Once
combined with the analytic contributions contained in the loop integrals \cite{Young:2002ib}, these
parameters form the renormalized low-energy coefficients of the chiral expansion.  Complete details
of the renormalization procedure are provided in the Appendix of Ref.~\cite{Young:2002ib}.

The loop-integral contributions $\beta^{\pi N}\!\left(m_\pi^2\right)$ and $\beta^{\pi
  \Delta}\!\left(m_\pi^2\right)$ are evaluated in the the heavy-baryon limit \cite{Jenkins:1990jv}
appropriate to a low-energy expansion.  The three-dimensional integral forms are
\cite{Hall:2013dva}
\begin{align}
\beta^{\pi N}\left(m_\pi^2\right) &= \frac{e^2}{4\,\pi}\,\frac{1}{288\,\pi^3\,f_\pi^2}\,\chi_N \int \dd^3k \, \frac{ \vec{k}^2\,u^2\left(k,\Lambda\right)}{\left( \vec{k}^2 + m_\pi^2\right)^3}, \label{eqn:betaPiN}\\
\beta^{\pi \Delta}\left(m_\pi^2\right) &= \frac{e^2}{4\,\pi}\,\frac{1}{288\,\pi^3\,f_\pi^2}\,\chi_\Delta \int \dd^3\,k \,u^2\left(k,\Lambda\right)\, \cross \nonumber \\ &\, \frac{ \omk^2\,\Delta\,\left(3\,\omk+\Delta\right) + \vec{k}^2\,\left(8\,\omk^2+9\,\omk\,\Delta + 3\,\Delta^2\right)}{8\,\omk^5\,\left(\omk+\Delta\right)^3}.
\label{eqn:betaPiDelta}
\end{align}
Here $\omk = \sqrt{\vec{k}^2 + m_\pi^2}$ is the energy carried by the pion which has three-momentum
$\vec{k}$, $\Delta$ is the aforementioned mass splitting between the Delta baryon and the nucleon,
$\Delta \equiv M_\Delta - M_N = 292$ MeV, and the pion decay constant is taken as $f_\pi = 92.4$
MeV.  The dipole regulator
\begin{align}
u\left(k,\Lambda\right) = \frac{1}{\left ( 1 + \vec{k}^2 / \Lambda^2\right)^{2}} \,,
\end{align}
of \eqnrtwo{eqn:betaPiN}{eqn:betaPiDelta} ensures that only soft momenta flow through
the effective-field theory degrees of freedom.

% Partial quenching
The lattice QCD results do not incorporate contributions from photons coupling to the disconnected
sea-quark loops of the vacuum which form the full meson dressings of $\chi$EFT -- they are
electro-quenched.  Thus, it is necessary to model the corrections associated with these effects.
This is done using partially quenched $\chi$EFT.  In this case, the standard coefficients for full
QCD
\begin{align}
\chi_N &= 2\,g_A^2,
\label{eqn:chi_N} \\
\chi_\Delta &= \frac{16}{9}\,\mathcal{C}^2, 
\label{eqn:chi_Delta}
\end{align}
are modified to account for partial quenching effects \cite{Detmold:2006vu} as explained in
Ref.~\cite{Hall:2013dva}. Thus, when fitting the lattice QCD results, we use coefficients that
reflect the absence of disconnected sea-quark-loop contributions.
\begin{align}
\label{eqn:pQN}
\chi_N \to \chi_N^{pQ} &= 2 g_A^2 -(D-F)^2 - \frac{7}{27}(D+3F)^2,\\
\chi_\Delta \to \chi_\Delta^{pQ}&=\frac{16}{9}\mathcal{C}^2 -
\frac{2}{9}\mathcal{C}^2 \, .
\label{eqn:pQD}
\end{align} 
We use the standard values of $g_A = 1.267$ and $\mathcal{C} = -1.52$ with $g_A = D+F$ and the
$\SU(6)$ symmetry relation $F = \frac{2}{3}D$.

% Unquenching
In anticipation of accounting for the missing disconnected
sea-quark-loop contributions in the lattice QCD calculations, the value $\Lambda = 0.80$ GeV is
adopted \cite{Wang:2008vb,Young:2002cj,Leinweber:2004tc,Leinweber:2006ug,Wang:1900ta}. This
regulator mass defines a pion cloud contribution to masses~\cite{Young:2002cj}, magnetic
moments~\cite{Leinweber:2004tc}, and charge radii~\cite{Wang:2008vb}, which enables corrections to
the pion cloud contributions associated with missing disconnected sea-quark-loop contributions.
This particular choice of regulator mass defines a neutron core contribution insensitive to
sea-quark-loop contributions~\cite{Wang:2013cfp}.

% Finite volume effects
Finite-volume effects are considered by replacing the continuum integrals of the chiral expansion
with sums over the momenta available on the periodic lattice. We note that the lattice volume
varies slightly across the four lattice data points available due to our use of the Sommer scale.

\subsection{Analysis}

We proceed by calculating the integrals of \eqnrtwo{eqn:betaPiN}{eqn:betaPiDelta} in the finite
volume of the periodic lattice by replacing the continuum integrals of the chiral expansion with
sums over the momenta available.  As the lattice QCD results do not include the contributions of
disconnected sea-quark-loop contributions, the coefficients of \eqnrtwo{eqn:pQN}{eqn:pQD} are used in \eqnrtwo{eqn:betaPiN}{eqn:betaPiDelta}.  This calculation is carried
out at each quark mass considered on the lattice.

\begin{figure}[t!]
	\centering
	\includegraphics[width=\columnwidth,origin=c]{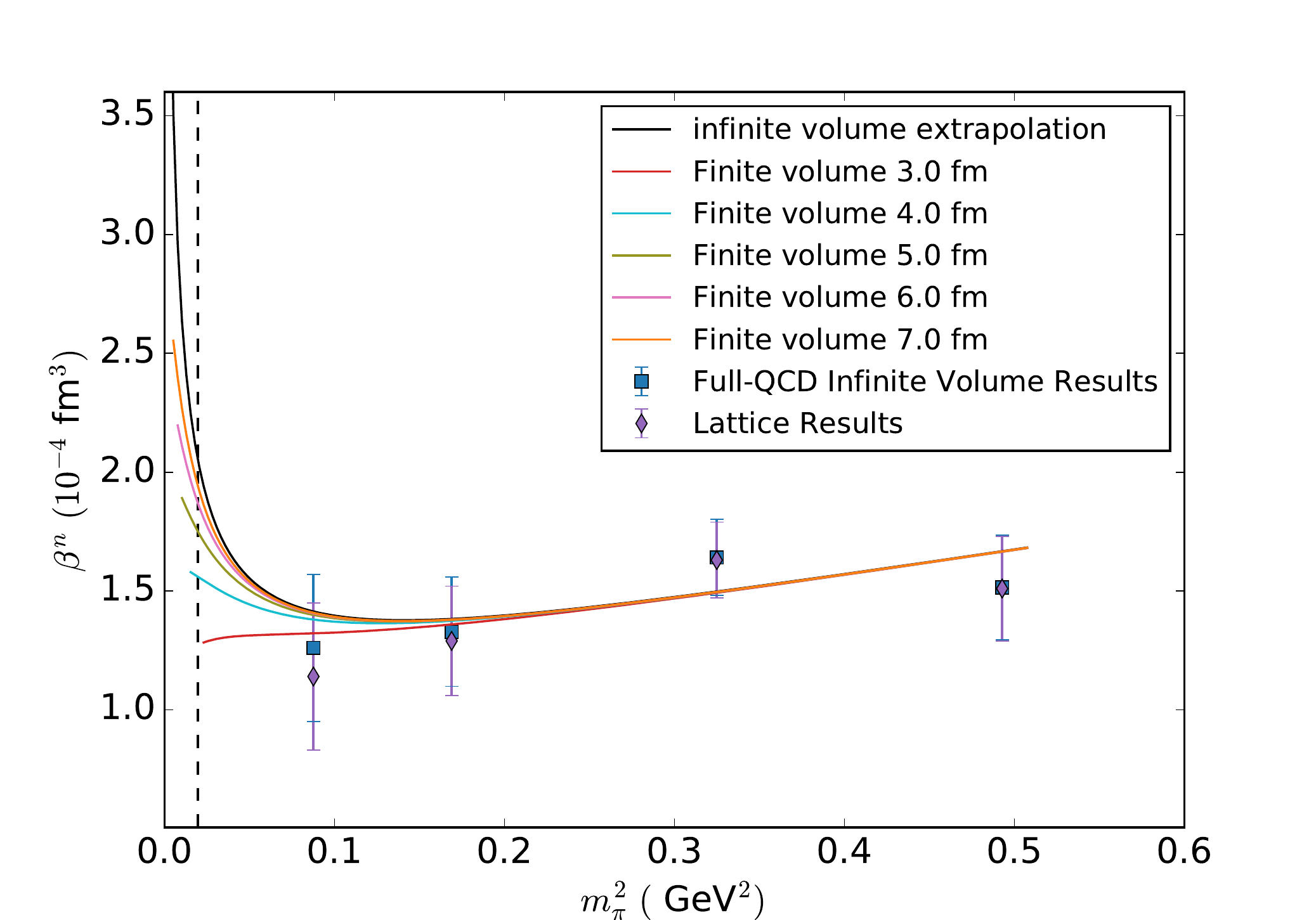}
	\caption{Correction of the lattice QCD results (violet diamond) for the neutron magnetic
          polarizability $\beta^n$ to infinite volume and full QCD (blue square) as
          described in the text.  Extrapolations of $\beta^n$ for a variety of spatial lattice
          volumes provide a guide to future lattice QCD simulations.  The infinite-volume case
          relevant to experiment is also illustrated.}
	\label{fig:chiEFT_2}
\end{figure}

One then numerically integrates \eqnrtwo{eqn:betaPiN}{eqn:betaPiDelta} in infinite volume and with
the full QCD coefficients of \eqnrtwo{eqn:chi_N}{eqn:chi_Delta}.  The difference between this
infinite-volume full-QCD result and aforementioned finite-volume partially quenched result at each
quark mass is used to correct the lattice QCD results to infinite volume and full QCD.  In this
way, both finite-volume and sea-quark-loop contribution corrections are incorporated. These
corrections are illustrated in \Fig{fig:chiEFT_2} by the (blue square) "Full-QCD
Infinite-Volume Results" next to the original (violet diamond) "Lattice Results."

At this point, the fit function of \eqnr{eqn:chiexpan} is fit to the corrected lattice QCD results
by adjusting the residual-series coefficients, $a_0$ and $a_2$.  Once $a_0$ and $a_2$ are
constrained, any volume can be considered.
Figure \ref{fig:chiEFT_2} shows chiral extrapolations for a range of volumes to provide guidance to
future lattice QCD simulations. Large box sizes are required in order to obtain an extrapolation
close to the infinite-volume value at the physical point. 

The physical polarizability is obtained from the constrained fit function of \eqnr{eqn:chiexpan}
with $m_\pi=m_\pi^{phys} = 140$ MeV. 
While the coefficients of the leading non-analytic terms of the chiral expansion are determined in a
model-independent manner, uncertainty in the higher-order terms of the expansion can be examined
through a variation of the regulator parameter $\Lambda$, which affects the sum of these
contributions.  Consideration of the broad range of $0.6 \leq \Lambda \leq 1.0$ GeV provides a systematic
uncertainty of $0.19 \times 10^{-3}$ fm$^3$ at the physical point.
Thus, we find \hbox{$\beta^n = 2.05(25)(19) \cross 10^{-4}$ fm$^3$} at the physical point. The
uncertainties are derived from the statistical errors of the fit parameters and the systematic
uncertainty associated with the chiral extrapolation respectively.

\begin{figure}[t!]
	\centering
	\includegraphics[width=\columnwidth,origin=c]{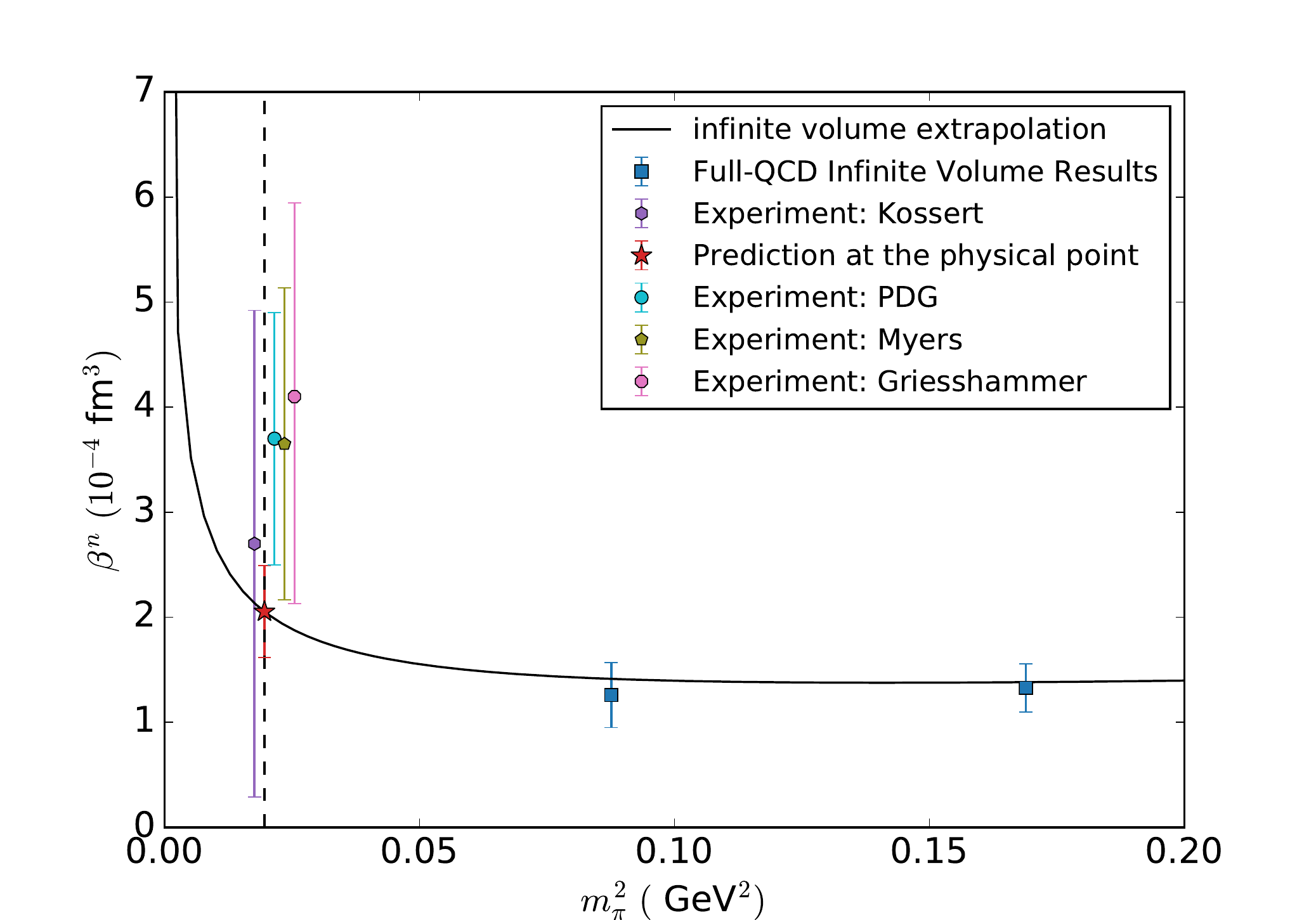}
	\caption{The magnetic polarizability of the neutron,
          $\beta_n$ obtained herein is compared with experimental results. The
          uncertainties in the lattice results contain both statistical and
          systematic errors simply added together. This is a conservative approach to produce a reliable estimation. Experimental results
          from Griesshammer
          \textit{et\ al.}~\cite{Griesshammer:2012we}, the
          PDG~\cite{Patrignani:2016xqp}, Kossert
          \textit{et\ al.}~\cite{Kossert:2002jc,Kossert:2002ws} and
          Myers \textit{et\ al.}~\cite{Myers:2014ace} are offset for
          clarity. }
	\label{fig:chiEFT_1}
\end{figure}

A comparison between this result and the experimental data is provided in
\Fig{fig:chiEFT_1}. Our calculation is in good agreement with a number of the experimental
results and poses an interesting challenge for greater experimental precision. Similarly, progress
in experimental measurement would drive further lattice QCD and chiral effective-field theory work.

These lattice results use a single lattice spacing and as such, it is not possible to quantify an
uncertainty associated with taking the continuum limit. However as a non-perturbatively improved
clover fermion action is used, the $\order{a^2}$ corrections are expected to be small relative to
the uncertainties already presented. It is anticipated that there is some degree of additive quark
mass renormalization due to the interaction of the background field with the Wilson term in the
fermion action~\cite{Bali:2017ian}, and the extent to which this small effect remains with the
clover fermion action is under investigation~\cite{Bignell:inprep}.

\section{Conclusion}

The neutron magnetic polarizability has been calculated using a novel approach in which asymmetric operators are used at the source and sink. The use of gauge-invariant Gaussian smearing at the source encapsulates the dominant QCD dynamics, while a gauge-fixed $U(1)$ two-dimensional eigenmode projection technique is used at the sink to encode the Landau level physics resulting from the presence of the uniform magnetic field. A systematic exploration of the parameter space was used to optimize operators that couple efficiently to the neutron ground state in a magnetic field. The use of this Landau mode projection at the sink has for the first time enabled the fitting of plateaus in the magnetic polarizability energy shifts.  

\par
Calculations at several pion masses have enabled the use of heavy baryon-chiral effective-field theory to relate the lattice QCD results to experiment. This enables us to make a theoretical prediction for the neutron magnetic polarizability of $\beta^n = 2.05(25)(19) \cross 10^{-4}$ fm$^3.$  
This prediction is founded on \emph{ab initio} lattice QCD simulations, with effective-field theory techniques used to account for the finite volume of the lattice, disconnected sea-quark-loop contributions, and extrapolation to the light quark masses of nature. The resulting value is in agreement with the current experimental estimates and presents an interesting challenge for greater experimental precision.
\par
A range of finite-volume extrapolations is performed in the framework of chiral effective-field theory. As these curves incorporate the contributions of sea-quark loops, they are presented as a guide to future lattice QCD simulations. At the physical pion mass, the $7$ fm curve still differs from the infinite-volume prediction by 6$\%$.
\par
Our result does not directly incorporate the sea-quark-loop effects of the magnetic field in the lattice simulation, which would require a separate Monte Carlo ensemble for each value of $B$ considered and as such is prohibitively expensive. We also note that in this study we have not considered the effects from the $B$-dependent additive quark mass renormalization that arises due to the use of Wilson-type fermions~\cite{Bali:2017ian}. The detailed impact of this effect on clover fermions has not been studied and will be addressed in a separate work that is in preparation \cite{Bignell:inprep}. While the effect can be observed in pion correlators, our preliminary tests indicate it is hidden by the large statistical fluctuations inherent in baryon correlators.
\par
It is interesting to consider that, in obtaining the magnetic polarizability, we want to work with small $\vec{B}$-field strengths in order to make use of the perturbative energy expansion for the neutron. This means we are the confining phase of QCD, such that quarks cannot have individual Landau levels. Nonetheless, the success of our Landau mode sinks indicates that the effects of the magnetic field on the quark distribution in the neutron are significant.
\par
Future work will examine the effect of using a gauge-covariant sink projection based on the eigenmodes of the two-dimensional $U(1)\times SU(3)$ Laplacian~\cite{Kamleh:2017yjx}. This alters the Landau level structure, breaking the degeneracy and mixing different levels. Indeed, a recent finite temperature study using the staggered quark formulation found that the contribution of the lowest Landau level eigenmodes remains important even after QCD interactions are introduced~\cite{Bruckmann:2017pft}, further motivating our investigation of the effectiveness of an eigenmode-projected sink which is aware of the QCD gauge field.

Another potential avenue for future investigation would be to explore relativistic
  corrections to the energy-field expansion of \eqnr{eqn:n:EB}.  To
  move beyond the use of Eq.~(\ref{eqn:n:EB}) requires one to separate
  the ratio of correlators in Eq.~(\ref{eqn:RBt}) to fit spin-field
  aligned and anti-aligned correlators separately.  It will be
  interesting to examine the behavior of these correlators and the
  extent to which QCD correlations can be exploited to obtain accurate
  fits for the energies $E(B) + m$ and $E(B) - m$.  These
  considerations may be particularly important in the study of
  charged hadrons. \par

\section*{Acknowledgments}
We thank the PACS-CS Collaboration for making their $2+1$ flavour configurations available and the ongoing support of the International Lattice Data Grid (ILDG). This work was supported with supercomputing resources provided by the Phoenix HPC service at the University of Adelaide. This research was undertaken with the assistance of resources from the National Computational Infrastructure (NCI). NCI resources were provided through the National Computational Merit Allocation Scheme, supported by the Australian Government and the University of Adelaide Partner Share. R.B. was supported by an Australian Government Research Training Program Scholarship. This research is supported by the Australian Research Council through Grants No.~DP150103164 and No.~DP120104627 (D.B.L).

%\bibliography{base}

\end{document}